\documentclass[prb,twocolumn,showpacs,superscriptaddress,preprintnumbers,amssymb]{revtex4-2}
\usepackage{graphicx}
\usepackage{dcolumn}
\usepackage{bm}
\usepackage[nointegrals]{wasysym}
\usepackage{color}
\usepackage{physics}
\usepackage{hyperref}

\newcommand{\beq}{\begin{equation}}
\newcommand{\eeq}{\end{equation}}
\newcommand{\beqn}{\begin{eqnarray}}
\newcommand{\eeqn}{\end{eqnarray}}

\newcommand{\ra}{\rightarrow}

\newcommand{\cF}{ {\cal F} }

\newcommand{\cH}{ {\cal H} }

\newcommand{\cL}{ {\cal L} }

\newcommand{\cS}{ {\cal S} }

\newcommand{\cZ}{ {\cal Z} }
\newcommand{\vect}[1]{{\bm{#1}}}
\newcommand{\Int}{\mathrm{int}}
\newcommand{\ii}{\mathrm{i}}

\newcommand{\hrho}{\hat{\rho}}

\newcommand{\hO}{\hat{O}}

\renewcommand{\O}{\mathrm{O}}
\newcommand{\SU}{\mathrm{SU}}
\newcommand{\U}{\mathrm{U}}
\newcommand{\Sp}{\mathrm{Sp}}

\newcommand{\nmj}[1]{{\color{black} #1}}
\newcommand{\cx}[1]{{\color{black} #1}}

\begin{document}

\title{Conformal Field Theories generated by Chern Insulators \\ under Quantum Decoherence}

\author{Kaixiang Su}

\author{Nayan Myerson-Jain}

\author{Cenke Xu}

\affiliation{Department of Physics, University of California,
Santa Barbara, CA 93106}


\begin{abstract}

We demonstrate that the fidelity between a pure state trivial insulator and the mixed state density matrix of a Chern insulator under decoherence can be mapped to a variety of two-dimensional conformal field theories (CFT); more specifically, the quantity $\cZ = \tr\{ \hrho^D_c \hrho_\Omega \}$ is mapped to the partition function of the desired CFT, where $\hrho^D_c$ and $\hrho_\Omega$ are respectively the density matrices of the decohered Chern insulator and a pure state trivial insulator. For a pure state Chern insulator with Chern number $2N$, the fidelity $\cZ$ is mapped to the partition function of the $\U(2N)_1$ CFT; under weak decoherence, the Chern insulator density matrix can experience certain instability, and the ``partition function" $\cZ$ can flow to other interacting CFTs with smaller central charges. The R\'{e}nyi relative entropy $\cF = - \log \tr\{ \hrho^D_c \hrho_\Omega \}$ is mapped to the free energy of the CFT, and we demonstrate that the central charge of the CFT can be extracted from the finite size scaling of $\cF$, analogous to the well-known finite size scaling of $2d$ CFT~\cite{cardyfinite, affleckfinite}. 

\end{abstract}

\maketitle

\section{Introduction}

The $2d$ Chern insulator~\cite{haldanechern} is the archetypal example of a topological insulator (TI). 
The Chern insulator is defined as the ground state (hence a pure state) of a two dimensional tight binding Hamiltonian of electrons. In reality, any quantum system is exposed to the environment, and experience certain level of decoherence through forming entanglement with the ancilla degrees of freedom in the environment. In the last few years TIs or more generally symmetry protected topological (SPT) states in open systems have attracted great interests, and these studies belong to a larger paradigm of classifying and characterizing topological features of density matrices ~\cite{TO_thermal, DMCI, UlhmannViyuela, TI_open1, TI_open2, de_Groot_2022, mawang, sptdecohere, bistrange,TI_dephase}. The strongest kind of decoherence would be thermalization, when the system reaches a thermal equilibrium with the environment after a long time evolution (compared with the microscopic time scales of the system) by interacting and entangling with the ancilla degrees of freedom in the environment. Under this type of long time and massive scale of decoherence, many of the topological features of the Chern insulator (such as its nontrivial edge states) would be lost, and the topological insulators are strictly well-defined at zero temperature.  

But one can also consider much weaker decoherence caused by exposing the system to the environment for a short amount of time, and the system will form weak entanglements with the ancilla qubits. In this case, after tracing out the ancilla qubits, the pure state topological insulator will still be rendered a mixed density matrix, but in stark contrast with thermalization, it is possible that some of the topological features survive this procedure. This type of weak decoherence is equivalent to the system being ``weakly measured" by the environment, but there is no ``post-selection" on the measurement outcomes. In the past few years phases and phase transitions that involve quantum measurements have been actively pursued, both theoretically and experimentally~\cite{nahummeasure,fishermeasure,PurificationTransion2020,ChoiBaoQiEhud2020,jianmeasure,BaoChoiEhud2020,GullansHuseProbes,SangHsieh2021,barkeshlimeasure,Ippoliti2021,vijaymeasure,reviewmeasure,nat1,nat2,nat3,natnishimori,leenishimori,nat4,garrattmeasure,googlemeasure,MIPTExpSC,MIPTTrapIon}. The fate of the mixed density matrix after decoherence strongly depends on the symmetry of the decoherence channel, as was pointed out in recent works. It was shown recently that the topological information of a mixed state density matrix can be extracted through ``strange correlators", which were originally devised to probe the pure state SPT states and TIs~\cite{YouXu2013,sptdecohere,bistrange}. In particular, a notion of type-II strange correlator and a formalism in terms of a ``doubled Hilbert space" was developed in order to extract the full anomaly of the mixed density matrix~\cite{sptdecohere}.  

The notion of the strange correlator is based on the overlap between the SPT state and a trivial state. Motivated by the recent progresses on understanding the topological features of mixed state density matrices, in this work we investigate the fidelity between a decohered Chern insulator density matrix and a pure state trivial insulator. We demonstrate that the fidelity can be mapped to the partition functions of a series of conformal field theories (CFT), whose central charges depend on the decoherence channel. We also propose a method to extract the central charge of the effective CFT, which is analogous to the finite size scaling of the actual free energy of ordinary CFTs~\cite{cardyfinite, affleckfinite}. This analysis enables us to extract the information of mixed state Chern insulator density matrix without having to know exactly which order parameter to use for the strange correlator.

\section{Chern insulator with $\nu = 1$}

\begin{center}
\begin{figure}
\includegraphics[width=0.47\textwidth]{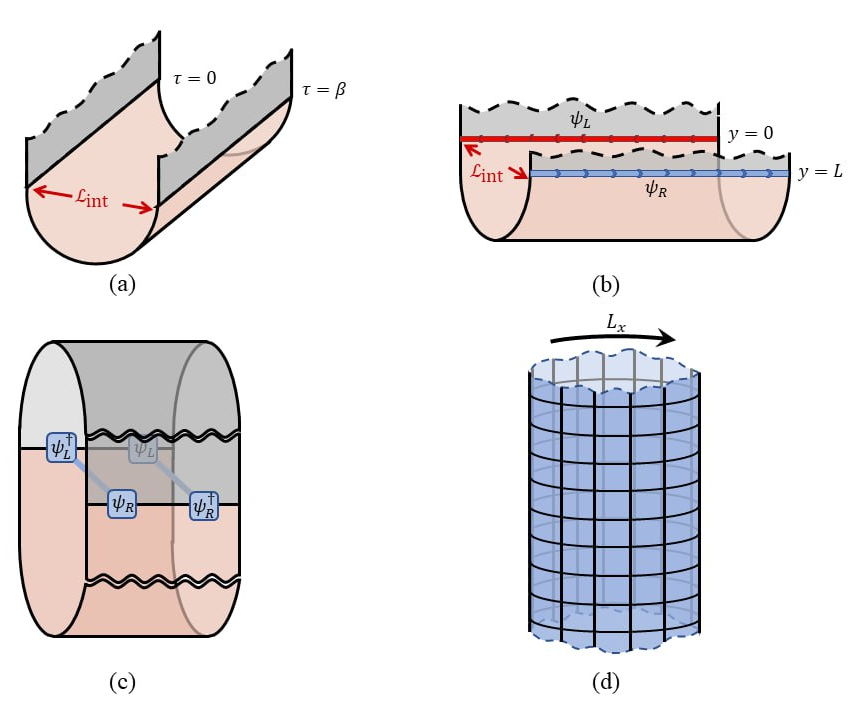}
\caption{(a) The fidelity $\cZ$ corresponds to path integral in Euclidean space-time, with interfaces at $\tau = 0$ and $\tau = \beta$; the decoherence is mapped to interaction between the two interfaces. (b) Under space-time rotation, the temporal interfaces at $\tau = 0,\beta$ become spatial interfaces $y = 0, L$, and there are chiral edge modes at each interface. (c) Example of type-II strange correlator. (d) The central charge of the effective CFT can be extracted from the finite size scaling of the R\'{e}nyi relative entropy $\cF$ on a cylindrical geometry. } \label{fig1}
\end{figure}
\end{center}

We are most interested in the following quantities: \beqn \cZ = \tr \{ \hrho^D_c \hrho_\Omega \}, \ \ \ \cF = - \log \cZ; \label{fidelity}\eeqn The density matrix of a pure state Chern insulator is $\hrho_c = |\Psi_c \rangle \langle \Psi_c |$, and $\hrho_\Omega$ is the density matrix of a trivial insulator. The quantity $\cZ$ is the fidelity between the Chern insulator density matrix (with or without decoherence) and a pure state trivial insulator density matrix, and \cx{$\cF$ is the R\'{e}nyi relative entropy}. The definition of the fidelity between two general mixed density matrices is more complicated, but it reduces to the simple form of Eq.~\ref{fidelity} when one of the density matrices is pure. In our work, it suffices to always keep $\hrho_\Omega$ the density matrix of the pure state trivial insulator. Later we shall see that $\cZ$ is mapped to the partition function of an effective $(2+0)d$ or $(1+1)D$ {\it nonchiral} conformal field theory (CFT), and $\cF$ is mapped to the free energy of the CFT. \cx{If we compute the overlap between the wave functions (rather than density matrices) of a pure state Chern insulator and trivial insulator, the overlap would become the partition function of a chiral CFT~\cite{shankar}. } Throughout the paper we will always use $d$ to label the spatial dimension, and $D$ to label the space-time dimension. 

The quantities $\cZ$ and $\cF$ encode the ``distance" between the two density matrices. We start with a tight-binding Hamiltonian that realizes the Chern insulator with Chern number $\nu = 1$, but we would like to pursue generic physics that is independent of the microscopic details, hence we tune the system close to the topological transition to a trivial insulator. In the case where the system is tuned close to the topological transition, the difference between the Chern insulator and the trivial insulator can be well captured by a single Dirac fermion with a mass (Fermion doubling demands there be another Dirac fermion in the momentum space, with presumably a much larger mass), where the sign of the mass determines whether the system is a Chern insulator or a trivial insulator. We also assume that there is sufficient crystal symmetry such as spatial inversion and discrete rotation, which render terms that deviate from a simple Dirac fermion irrelevant in the infrared when the Dirac mass is small. One example of such model is the spinless version of the so-called BHZ model on the square lattice~\cite{BHZ}, which has multiple discrete symmetries such as charge-conjugation, spatial inversion, discrete rotation, etc. These symmetries ensure that close to the topological-trivial transition with band inversion, the system is well captured by a single Dirac fermion at the Gamma point of the Brillouin zone with emergent Lorentz symmetry. The Hamiltonian and the Euclidean space-time Lagrangian of the Dirac fermion read: \beqn H &=& \int d^2x \ \cH =  \int d^2x \ \psi^\dagger (\ii \sigma^z \partial_x + \ii \sigma^x \partial_y + m \sigma^y) \psi; \cr\cr \cL &=& \psi^\dagger (\partial_\tau + \ii \sigma^z \partial_x + \ii \sigma^x \partial_y + m \sigma^y) \psi \cr\cr &=& \bar{\psi} (\gamma_0 \partial_\tau + \gamma_1 \partial_x + \gamma_2 \partial_y)\psi + m \bar{\psi}\psi \cr\cr && (\gamma_0, \gamma_1, \gamma_2) = (\sigma^y, - \sigma^x, \sigma^z) \label{Dirac2d} \eeqn with $\bar{\psi} = \psi^\dagger \gamma_0$. 

We would like to consider a Chern insulator going through a decoherence quantum channel, where the ancilla qubits entangle with the system through weak interactions that preserve the total $\U(1)$ charge of the system. For example, the ancilla qubits can interact with the current density of the system. \cx{ After being exposed to the quantum channel, we need to trace out the ancilla qubits that couple to the current density (which corresponds to measuring the ancilla qubits and sum over all the measurement outcomes),} and the density matrix of the Chern insulator becomes a mixed density matrix which can take the following form \beqn \hrho^D_c &=& \otimes_{\vect{x}}\mathcal{E}_{\vect{x}}[\hrho_c],\cr\cr \mathcal{E}_{\vect{x}}[\hrho_c] &\sim& \hrho_c + \sum_{\mu} p_\mu \left( \hat{J}^\mu_\vect{x} \hrho_c \hat{J}^\mu_\vect{x} \right). \label{decohere} \eeqn
where $\hat{J}^\mu$ is the three component {\it Hermitian} current vector: \beqn \hat{J}^\mu &=& (\bar{\psi} \gamma^0 \psi, \ii \bar{\psi} \gamma^1 \psi, \ii \bar{\psi} \gamma^2 \psi ) \cr\cr &=& (\psi^\dagger \psi, - \psi^\dagger \sigma^z \psi, - \psi^\dagger \sigma^x \psi ). \eeqn Since the charge $\U(1)$ symmetry is a constraint on the quantum channel, the mixed density matrix after decoherence will be in the {\it canonical ensemble} with a fixed number of fermions. 


The exact density matrix of a system is often tedious to work with. For a quantum many-body system, we often only care about the long wavelength behaviors of the system, and this physics can be captured through coarse-graining or renormalization group techniques. There is a rather convenient formalism that allows us to use these techniques when we study the effects of decoherence. This formalism is based on the well-known fact that the density matrix of the ground state of a system can be generated through path integral in the Euclidean space-time $[\hrho_0]_{\phi_1(\vect{x}), \phi_2(\vect{x})} \sim \lim_{\beta \ra \infty } \int D \phi(\vect{x}, \tau) \exp \left( - \cS_0 \right) $, with the temporal boundary condition $ \phi(\vect{x},0) = \phi_1(\vect{x})$ and $\phi(\vect{x},\beta) = \phi_2(\vect{x})$. Here $\cS_0$ is the action of the system whose ground state is the desired pure state $|\Psi_0\rangle$. Then under decoherence, the mixed density matrix becomes~\cite{garrattmeasure,wfdecohere} (Fig.~\ref{fig1}(a)) \beqn && [\hrho^D]_{\phi_1(\vect{x}), \phi_2(\vect{x})} \sim \lim_{\beta \ra \infty } \int \ D \phi (\vect{x}, \tau) \exp\left( - \cS_0 - \cS^{\Int} \right); \cr\cr && \cS^{\Int} = \int d\vect{x} \ \cL^{\Int} ( \phi(\vect{x}, 0), \phi(\vect{x}, \beta) ), \label{pathintegral1} \eeqn the boundary condition $ \phi(\vect{x},0) = \phi_1(\vect{x})$, $\phi(\vect{x},\beta) = \phi_2(\vect{x})$ must hold in this path integral. The effect of decoreherene is mapped to the interaction $\cL^{\Int}$ between fields at $\tau = 0$ and $\tau = \beta$.

The symmetry is the most important condition one needs to consider in order to determine the form of $\cL^{\Int}$. A density matrix can have a {\it doubled} (or strong) symmetry condition when it is invariant under separate actions from left and right multiplication of the symmetry operation~\cite{de_Groot_2022,mawang,sptdecohere}. When the measurement from the quantum channel explicitly preserves the charge-$\U(1)$ symmetry, $\cL^{\Int}$ needs to preserve separate $\U(1)$ transformations acting from the left and right side of the density matrix, which translate to the separate U(1) symmetries of the fields at $\tau = 0$ and $\tau = \beta$. In this case $\cL^{\Int}$ should include the following term: \beqn \cL^{\Int} \sim \sum_\mu - g_\mu J_\mu(\vect{x},0) J_\mu(\vect{x},\beta) \eeqn where $g_\mu \sim p_\mu$ for small $p_\mu$. The charge current operator $\vect{J} = (J_x, J_y)$ is odd under spatial inversion, and all components of $J_\mu$ are odd under charge conjugation. \cx{ In addition to the doubled U(1) symmetry, we demand the decoherence channel preserve the {\it diagonal} (or weak) charge conjugation and spatial inversion symmetry, which physically means that all the measurement outcomes are summed over without bias.} This condition precludes terms linear with current operators $J_\mu(\vect{x},0)$, $J_\mu(\vect{x},\beta)$ in $\cL^{\Int}$.

Hence the quantity $\cZ$ can be computed through a path integral of a Chern insulator action sharing temporal interfaces with a trivial insulator at $\tau = 0, \beta$, and this computation can be approximated by a path integral with the following action on the $(2+0)d$ temporal interfaces $\tau = 0, \beta$ \beqn \cS = \int d^2x \ \bar{\psi}(\gamma_1 \partial_1 + \gamma_2 \partial_2) \psi + \cL^{\Int}, \eeqn \cx{as there are low energy modes localized at the temporal interfaces. The low energy modes at the temporal interface are eigensates of $\sigma^y$. } It is much more physically intuitive for us to perform a space-time rotation in the $(\tau, y)$ plane, and position the interface at $y = 0, L$, \cx{and the modes localized at the spatial interface would be eigenstates of $\sigma^z$.} The action that describes the low energy states at both boundaries is \beqn \cS_{1d} &=& \int d\tau dx \ \cL_{1d} = \int d\tau dx \ \psi^\dagger (\partial_\tau + \ii \sigma^z \partial_x)\psi \cr\cr &=& \int d\tau dx \ \bar{\psi} (\gamma_0 \partial_\tau + \gamma_1 \partial_x ) \psi. \label{Dirac1d} \eeqn This action corresponds to a $1d$ Hamiltonian of Dirac fermion, whose bosonized form is \beqn H_0 = \int dx \ \cH_0 (x), \ \ \cH_0(x) = \frac{K}{2\pi} (\nabla_x \theta)^2 + \frac{1}{2\pi K}(\nabla_x \phi)^2. \eeqn For free fermions without any interaction, $K = 1$. Here we stress that Eq.~\ref{Dirac1d} and its bosonized form already include physics at the two opposite spatial interfaces, which correspond to the temporal interfaces $\tau = 0$ and $\beta$ combined before the space-time rotation. 

After the space-time rotation, $\cL^{\Int}$ would take the form of interaction between currents on the spatial interfaces $y = 0$ and $y = L$: \beqn \cL^{\Int} \sim \sum_\mu - g_\mu J'_\mu(x,0) J'_\mu(x,L), \eeqn where \beqn J'_\mu &=& (\bar{\psi}\gamma_2\psi, \ii \bar{\psi}\gamma_1\psi, \ii \bar{\psi}\gamma_0\psi) \cr\cr &=& (- \ii \psi^\dagger \sigma^x \psi, - \psi^\dagger \sigma^z \psi, \ii \psi^\dagger \psi). \eeqn The $g_0$ term does not project to an obvious nontrivial operator at the spatial boundary after space-time rotation; $g_1$ and $g_2$ have very similar effect, as they both project to the following term at the spatial boundary (Fig.~\ref{fig1}(b)): \beqn \cL^{\Int} \sim \cH^{\Int} = g \rho_L(x) \rho_R(x), \ \ g \sim g_1 + g_2, \eeqn hence the decoherence is now mapped to a ``{\it repulsive}" interaction between the left and right moving charge densities in the effective $(1+1)D$ system corresponding to the Hamiltonian density $\cH_0 + \cH^{\Int}$.

This interaction would renormalize the Luttinger parameter $K$, i.e. $ K = \sqrt{(2\pi - g)/(2\pi + g)}$,  but it will not gap out the boundary states. This is because the ``doubled symmetry" condition demands that the effective action after space-time rotation be invariant under independent $\U(1)_L$ and $\U(1)_R$ symmetries which act on the left and right moving modes of $\cH_0$, and it is well-known that a system with these symmetries have perturbative 't Hooft anomaly and cannot be gapped.

\section{Chern insulator with $\nu = 2$}

We now consider two copies of Chern insulators described by the BHZ model both with Chern number $\nu = 1$, topologically this is also equivalent to a Chern insulator with Chern number $\nu = 2$. After going through the decoherence quantum channel which preserves the charge number, the density matrix of the Chern insulator becomes \beqn \hrho^D_c &=& \otimes_{\vect{x}}\mathcal{E}_{\vect{x}}[\hrho_c], \cr\cr \mathcal{E}_{\vect{x}}[\hrho_c] &\sim& \hrho_c + \sum_{i = 1,2} \sum_{\mu} p_{i,\mu} \left( \hat{J}^\mu_{i,\vect{x}} \hrho_c \hat{J}^\mu_{i,\vect{x}} \right) \cr\cr &+& p_e \ \left( \hO_\vect{x} \hrho_c \hO^\dagger_\vect{x} + \hO^\dagger_\vect{x} \hrho_c \hO_\vect{x} \right), \cr\cr \hO_\vect{x} &=& c^\dagger_{1,\vect{x}} \sigma^z c_{2,\vect{x}}, \label{densitymatrix2} \eeqn with small $p_{i,\mu}$ and $p_e$. The subscript $i = 1,2$ labels the two copies of the Chern insulators. In addition to interacting with the current operators of the system, the ancilla qubits can also interact with the exciton operator $\hO$, if the interaction with the ancilla qubits preserve the total charge $\U(1)$ symmetry. Our goal is to explore the possible effective CFTs that can be realized through $\tr\{ \hrho^D_c \hrho_\Omega \}$, hence a specific form of decoherence on exciton operators was chosen in Eq.~\ref{densitymatrix2}, which ensures that the realized CFT is unitary.

Following the procedure in the previous section, after a space-time rotation in the $(y,\tau)$ plane, the quantity $\cZ = \tr\{ \hrho^D_c \hrho_\Omega \}$ can be mapped to the following path integral of $(1+1)D$ interacting non-chiral fermions: \beqn \cZ &\sim& \int D[\psi_i]D[\psi^\dagger_i] \exp \left( - \int d\tau dx \ \cL_{1d} + \cL^{\Int} \right) \cr\cr \cL^{\Int} &=& \sum_{i = 1,2} g_{i} \rho_{i,L} \rho_{i,R} + g_e ( \psi^\dagger_{1,L}\psi_{2,L}\psi^\dagger_{2,R}\psi_{1,R} + h.c.), \label{pathintegral2} \eeqn where $\cL_{1d}$ is the Lagrangian of two flavors of free $(1+1)D$ Dirac fermion. We can turn the path integral problem Eq.~\ref{pathintegral2} into a Hamiltonian formalism in $1d$: \beqn H &=& \int dx \ \sum_{i = 1,2} \frac{K_i}{2 \pi} (\nabla_x \theta_i) + \frac{1}{2 \pi K_i} (\nabla_x \phi_i)^2 \cr\cr &+& u \cos( 2\phi_1 - 2 \phi_2). \eeqn Here $u$ is proportional to $g_e \sim p_e$, and $ K_i = \sqrt{(2\pi - g_i)/(2\pi + g_i)}$  with $ g_i \sim p_{i,1} + p_{i,2} > 0$. Here $\exp(\ii 2 \phi_i) \sim \psi^\dagger_{i,L} \psi_{i,R}$ corresponds to the back scattering terms, and $\exp(\ii 2 \theta_i) \sim \psi_{i,L} \psi_{i,R}$ corresponds to the Cooper pairs.

For the most natural choice of parameters where $K_i < 1$, the vertex operator $ u \cos(2\phi_1 - 2\phi_2)$ is {\it relevant}. A relevant $u$ will gap out the channel $\theta_- \sim \theta_1 - \theta_2$ and $\phi_- \sim \phi_1 - \phi_2$, while leaving the channel $\theta_+ \sim \theta_1 + \theta_2$ and $\phi_+ \sim \phi_1+\phi_2$ gapless and algebraic. All the fermions will acquire a gap, for example \beqn \psi_{1,L} \sim e^{\ii \theta_1 + \ii \phi_1 } \sim e^{ \ii \frac{1}{2}(\theta_+ + \theta_-) + \ii \frac{1}{2} (\phi_+ + \phi_-) }, \eeqn as the short range correlation of the $\phi_-$ channel will render all the fermion operators short-ranged. But some composite operators will acquire quasi-long range or power-law correlation, such as the following four body operators \beqn \psi_{1,L}\psi_{1,R} \psi_{2,L}\psi_{2,R} &\sim& e^{\ii 2 \theta_+} . \eeqn The correlation function $\langle e^{\ii 2 \theta_+(0,0)} e^{- \ii 2 \theta_+(0,x)} \rangle$ of the $(1+1)D$ CFT is actually a ``type-II strange correlator" that was proposed in Ref.~\onlinecite{sptdecohere} in the $(2+0)d$ space before the space-time rotation: \beqn && \langle e^{\ii 2 \theta_+(0, 0)} e^{- \ii 2 \theta_+(\tau, x)} \rangle \cr\cr &\sim& \tr\{ \hrho^D_c \hat{\Delta}(0) \hat{\Delta}^\dagger(\vect{x}) \hrho_\Omega \hat{\Delta} (0) \hat{\Delta}^\dagger(\vect{x})   \} \eeqn where $\hat{\Delta}(x)$ is a Cooper pair operator $\hat{\Delta}(\vect{x}) \sim c_{1,\alpha}(\vect{x}) c_{2,\alpha}(\vect{x})$ with the same Dirac index $\alpha = 1,2$, and $\vect{x} = (x, y)$ labels the $2d$ spatial coordinate before the space-time rotation

Hence even under weak decoherence, the Chern insulator experiences certain ``{\it instability}" in the sense that the CFT whose partition function corresponds to $\tr\{ \hrho^D_c \hrho_\Omega \}$ can have its central charge reduced from $c = 2$ to $c = 1$ even under weak decoherence. The change of the central charge can be extracted through the finite size scaling of $\cF = - \log \cZ$, which we will discuss later in this manuscript. 

\section{Chern insulator with $\nu = 2N$}

Now we consider $\nu = 2N$ copies of the Chern insulator. The system can have a large $\U(2N)$ flavor symmetry. \cx{ There could be many possible decoherence channels on the system. As an example, we choose the decoherence channel that acts on the Chern insulator density matrix in the following way:} \beqn \mathcal{E}_{\vect{x}}[\hrho_c] &\sim& \hrho_c + \sum_{a} \sum_{\mu} p_{a,\mu} \left( \hat{J}^{a, \mu}_{\vect{x}} \hrho_c \hat{J}^{a, \mu}_{\vect{x}} \right), \cr\cr \hat{J}^{a, \mu}_{\vect{x}} &=& (\bar{\psi} \gamma^0 \tau^a \psi, \ii \bar{\psi} \gamma^1 \tau^a \psi, \ii \bar{\psi} \gamma^2 \tau^a \psi ), \eeqn $\hat{J}^{a, \mu}_{\vect{x}}$ with $a = 1,2,3$ is the current operator for the $\SU(2)$ subgroup of the $\U(2N)$ flavor symmetry. This decoherence channel corresponds to the physics that the ancilla qubits of the quantum channel interact with the currents of the $\SU(2)$ flavor symmetry, and eventually the ancilla qubits are traced out. We would like to keep a {\it diagonal} $\SU(2)$ flavor symmetry, hence $p_{a,\mu}$ does not depend on the $\SU(2)$ index $a$. 

Following the same procedure as before, after space-time rotation we arrive at the following effective Hamiltonian at the $(1+1)D$ space-time: \beqn \cH(x) = \cH_0(x) + \sum_a g J^a_L(x)  \cdot J^a_R(x). \eeqn Since the decoherence channel breaks the doubled $\SU(2)$ symmetry down to the diagonal $\SU(2)$, the $\sum_a g J^a_L(x) \cdot J^a_R(x)$ term is allowed, and most naturally $g > 0$. This interaction is {\it marginally relevant}, and its effect is to gap out the SU(2) sector of the CFT. The original $\U(2N)_1$ CFT has the following decomposition \cite{affleckU(N)}: \beqn && \U(2N)_1 \simeq \SU(N)_2 \oplus \U(1)_{2N} \oplus \SU(2)_{N}; \ \mathrm{or} \cr\cr && \U(2N) \simeq \O(4N)_1 \simeq \Sp(N)_1 \oplus \SU(2)_N. \eeqn Since the $\SU(2)_N$ sector of the CFT is gapped out by decoherence, eventually the central charge is reduced from $c = 2N$ to \beqn c = \frac{N(2N+1)}{N+2} .\eeqn Other CFTs with a coset construction can also be engineered through different types of decoherence. 

\section{central
charge of the effective CFT}

In previous sections we have shown that under decoherence the effective CFT with partition function $\cZ = \tr\{ \hrho_c\hrho_\Omega \} $ can become instable and flow to CFTs with smaller central charges. In this section we propose that, if we design our tight binding Hamiltonian on a cylinder with finite circumference $L$ along the $x$ direction, while infinite length along $y$ (Fig.~\ref{fig1}(d)), the ``free energy" $\cF = - \log \cZ$ per unit length along the $y$ direction will have the following finite size scaling \beqn \frac{\cF}{L_y} = f_0 L - \frac{\pi c }{6 L} + \mathcal{O} \bigg (\frac{1}{L^2} \bigg ), \eeqn where $ c$ is the central charge. This equation is directly inspired by the well-known formula derived in Refs.~\onlinecite{cardyfinite} and ~\onlinecite{affleckfinite} for the actual $2d$ CFTs. 

In the following we compute $\cF$ for the pure state Chern insulator, and demonstrate that its finite size scaling does encode the correct central charge. In Eq.~\ref{Dirac2d}, for the Chern insulator we take its mass to be $m>0$ while for the trivial insulator we take its mass to be $-m$. The Hamiltonian densities are 
\begin{align}
    &\mathcal{H}_{c} \sim \psi^\dagger(k)( -k_x \sigma^z - k_y\sigma^x + m\sigma^y) \psi(k), \\
    &\mathcal{H}_{\Omega} \sim \psi^\dagger(k)(-k_x \sigma^z - k_y \sigma^x - m\sigma^y) \psi(k).
\end{align}
The Bloch Hamiltonians can be diagonalized in the momentum basis labeled by
\begin{equation}
    k_y\in \mathbb{R}; \ \ \  k_x  = \frac{2\pi (n - 1/2)}{L}, n\in \mathbb{Z}.
\end{equation}
Here we have chosen an anti-periodic boundary condition for the fermions along the $x$ direction, as one can show that the ground state wave function for the Chern insulator with an anti-periodic boundary condition has a lower energy compared with that with a periodic boundary condition. 

The fidelity $\cZ$ becomes a product of the overlap between Bloch wave functions at each momentum: 
\beqn
\cZ &=& |\braket{\Psi_\Omega}{\Psi_c}|^2 = \prod_{k}b(k), \cr\cr b(k) &=& \frac{k_x^2+k_y^2}{k_x^2+k_y^2+m^2}.
\eeqn
The ``free energy" $\cF$ per unit length in $y$ direction is
\begin{equation}
\frac{\cF}{L_y} = - \frac{\log \cZ}{L_y} = -\frac{1}{2\pi}\sum_{k_x}\int dk_y \log(b(k_x,k_y))
\end{equation}
The integral in $k_y$ can be performed explicitly, and yields
\begin{equation}
  \frac{\cF}{L_y} =  \frac{1}{2\pi}\sum_{k_x}(-2\pi \abs{k_x} + 2\pi \sqrt{k_x^2 + m^2}) \label{Regularizedsum}.
\end{equation}

We examine the term $ \sum_{k_x} - |k_x|$ first
\begin{equation}
   \sum_{k_x} - |k_x| = \sum_{n = -\infty}^{\infty} - \frac{2\pi}{L}|n - 1/2| = \sum_{n = 1}^{\infty} - \frac{2\pi}{L}(2n - 1) 
\end{equation}
In order to perform the sum of all odd integers, we need to use the Zeta function regularization: \beqn \sum_{n = 1}^{\infty} (2n - 1)^{-s} = (1 - 2^{-s})\zeta(s). \eeqn Plugging in $s = -1$ would give us $\sum_n (2n - 1) = 1/12$. Hence eventually we obtain the following result:
\begin{equation}
 \sum_{k_x} - |k_x|  =  -\frac{\pi c}{6L}
\end{equation}
where $c = 1$. This result is consistent with the well-known formula of finite size scaling of $2d$ CFTs given in Ref.~\cite{cardyfinite, affleckfinite}. 

However, there is also correction from finite mass $m$ which is attained by analyzing the second term of Eq.~\ref{Regularizedsum}. This term $\sum_{k_x} \sqrt{k_x^2+m^2}$ would physically correspond to the Casmir effect of a massive particle in $1d$. Intuitively a massive particle would not lead to any Casmir effect with large mass, but we will test this intuition as follows. This sum can be regularized using a special form of the Abel-Plana formula for half-integer sums \cite{Casimirbook}:
\beqn
\sum_{n =0}^{\infty} f_{n+\frac{1}{2}} = \int_0^\infty dt f(t) - \ii \int_0^\infty dt \frac{ [f(\ii t) - f(-\ii t)]}{1+e^{2 \pi t}}. \label{Abel}
\eeqn
Only the second integral contains a finite piece independent of the regularization scheme, and for a large $|m|$ the integral can be approximated as $- \sqrt{\frac{2|m|}{\pi L}} e^{-|m|L}$. Therefore, for large $\abs{m}$, \cx{the quantity $\cF/L_y$ with anti-periodic boundary condition reads}
\begin{align}
    &\frac{\mathcal{F}_{\mathrm{APBC}}}{L_y} \sim -\frac{\pi c}{6L} + \sqrt{\frac{2\abs{m}}{\pi L}}e^{-\abs{m}L},
\end{align}
with $c = 1$. As expected, the correction to $\cF$ arising from the second term of Eq.~\ref{Regularizedsum} decays rapidly in the large mass limit and one recovers the exact CFT scaling. 

For completeness, one can also compute the sum Eq.~\ref{Regularizedsum} for the more energetic case of periodic boundary conditions where the momentum in the $x$-direction now takes values of $k_x \in \frac{2\pi}{L} \mathbb{Z}$, which would yield the result 
\begin{align}
    &\frac{\mathcal{F}_{\mathrm{PBC}}}{L_y} \sim \frac{\pi}{3L} - \sqrt{\frac{2\abs{m}}{\pi L}}e^{-\abs{m}L}.
\end{align}

\cx{One can also use the general formula Eq.~\ref{Abel} to evaluate the fist sum of Eq.~\ref{Regularizedsum}. The first integral in Eq.~\ref{Abel} diverges in the UV, and it is proportional to $L$; the second integral leads to the desired result of $-\pi/(6L)$.}

\section{Chern insulator with $\nu = 4$}

If we start with the Chern insulator with $\nu = 4$, the physics should be qualitatively similar to the case with other Chern numbers, as long as the decoherence channels only performs measurements on quantities that preserve the $\U(1)$ charge symmetry. The $\U(1)$ charge symmetry will always become a doubled symmetry in the density matrix formalism, and map to the $\U(1)_L$ and $\U(1)_R$ symmetry after the space-time rotation. Then the perturbative 't Hooft anomaly will exclude the possibility of a gapped ``partition function" $\cZ \sim \tr\{ \hrho^D_{c} \hrho_\Omega \}$.

Let us now break the doubled $\U(1)$ symmetry down to a diagonal $\U(1)$ symmetry, meaning we allow the quantum channel to measure all bosonic operators, such as all fermion bilinear operators including Cooper pairs. Effectively the $1d$ system after space-time rotation has four left-moving and four right-moving complex fermions, and there is only one global (diagonal) $\U(1)$ charge symmetry, when there is no post-selection of measurement outcomes. There is also a separate fermion parity of left and right chiral fermions: $Z^L_2 \times Z_2^R$. The decoherence is mapped to interacting terms of the $1d$ system. According to the classification of interacting TIs, four copies of $(1+1)D$ nonchiral comlex fermions can be gapped out by four fermion interactions without breaking any symmetry or ground state degeneracy \cite{fidkowski1,fidkowski2,Qi2013,LevinGufSPT}. This mechanism also generalizes to other models and higher dimensions, and is in general referred to as symmetric mass generation (SMG)~\cite{SMGwang1,kevinQSH,shailesh1,shailesh2,shailesh3,simon1,SMGAshvin,SMGashvin2,SMGwang2,SMGYouReview,SMGYou2022,anna1,tong1,tong2}. But the interaction terms in our case have a more restrictive form, and need to be generated by decoherence. We have {\it not} found the corresponding Kraus operators that can generate the exact form of the interactions that have been definitely proven to cause SMG for the system. We leave the possibility of SMG caused solely by decoherence as an open question for future study. 

By considering proper Kraus operators and their Hermitian conjugates that respect fermion parity, one can indeed gap out the CFT described by $\cZ$ and $\cF$. However we note here that the $(1+1)D$ system so obtained from space-time rotation is not a completely trivial state, instead it would spontaneously break the $Z_2^L \times Z_2^R$ symmetry down to a diagonal fermion parity. Let us consider the following decoherence channcel as an example: 
\beqn \mathcal{E}_{\vect{x}}[\hrho_c] &\sim& \hrho_c + \sum_{i=1}^4 p_i \left ( \hat{O}_{i,\vect{x}} \hat{\rho}_c \hat{O}_{i,\vect{x},}^\dagger + \hat{O}_{i,\vect{x}}^\dagger \hat{\rho}_c \hat{O}_{i,\vect{x}} \right ) \cr\cr \hO_{1,\vect{x}} &=& c_{1,\vect{x}}^\dagger \sigma^z (c_{2,\vect{x}}^\dagger)^T  \text{ and } \hO_{i \neq 1, \vect{x}} = c_{1,\vect{x}}^\dagger \sigma^z c_{i,\vect{x}}.
\eeqn 
Following our previous procedure, the fidelity is mapped to a path integral problem of four flavors of $(1+1)D$ Dirac fermions with the following interactions
\begin{equation}
\begin{aligned}
    \mathcal{L}^{\mathrm{int}} &= \sum_{i\neq 1} g_i (\psi_{1,L}^\dagger\psi_{i,L}\psi_{i,R}^\dagger\psi_{1,R}+\text{h.c.})\\&-g_1 (\psi_{1,L}^\dagger\psi_{2,L}^\dagger\psi_{2,R}\psi_{1,R} + \text{h.c.})
\end{aligned}
\end{equation}
These terms after Abelian bosonization become:
\beqn
    \mathcal{L}^{\mathrm{int}} &=& \sum_{i \neq 1} u_i \cos(2\phi_1 - 2\phi_i) - u_1 \cos(2\phi_1+2\phi_2) \cr\cr
    &=& \sum_{i\neq 1} u_i \cos(\Lambda^T_i \mathcal{K} \Phi ) - u_1 \cos(\Lambda^T_1 \mathcal{K} \Phi).
\eeqn
The vector $\Phi$ is $\Phi = (\phi_{L,1},\dots,\phi_{L,4};\phi_{R,1},\dots \phi_{R,4})^T$ where $\phi_{L,i} = \theta_i - \phi_i$ and $\phi_{R,i} = \theta_i + \phi_i$, and the $K$-matrix is $\mathcal{K} = \text{diag}(-1,\dots,-1;1,\dots,1)$. In this basis, one can verify that the four $\Lambda$-vectors take the following form
\beqn
\Lambda_1 &=& (1,-1,0,0;1,-1,0,0)^T \cr\cr
\Lambda_2 &=& (1,0,-1,0;1,0,-1,0)^T \cr\cr
\Lambda_3 &=& (1,0,0,-1;1,0,0,-1)^T \cr\cr
\Lambda_4 &=& (1,1,0,0;1,1,0,0)^T. \label{vector}
\eeqn
We will outline the conditions that these vectors satisfy and show that they drive the theory into a $Z_2$ SSB state. A more complete discussion of these conditions and the $\Lambda$-vector formalism can be found in Refs \cite{LevinGufSPT,LevinSternSPT}.

In order for these vertex operators to completely gap the system, the $\Lambda$ vectors must be four linearly independent vectors that satisfy the so-called Haldane null-vector condition: $\Lambda_i^T \mathcal{K} \Lambda_j = 0$ for all $i,j$. This condition ensures that the fields can be rotated to a new basis where each vertex operator contains only one single field instead of a linear combination, and this condition is indeed met by the four vectors in Eq.~\ref{vector}. \cx{However, if these terms can indeed trivially gap out the $(1+1)D$ system without any degeneracy (i.e. SMG), it is necessary that the determinants of all possible $ 4 \times 4$ minors of the matrix with rows of $\Lambda_i^T$ do not share a common factor larger than one.} One can verify that this condition is {\it not} met by the four interaction terms generated by Kraus operators chosen above, and our choice of Kraus operators must drive the boundary into a state that spontaneously breaks the fermion parity $Z_2^L \times Z_2^R$ down to its diagonal. As a consequence, the following fermion bilinears in the effective $(1+1)D$ theory acquire a nonzero expectation value 
\beqn
\langle \psi_{L,i }^\dagger \psi_{R,i  } \rangle \neq 0 .
\eeqn
The correlation of these fermion bilinears are the following type-II strange correlator first introduced in Ref.~\cite{sptdecohere} (Fig.~\ref{fig1}(c)):
\beqn
&& \langle \psi_{L,i }^\dagger \psi_{R,i  } (0, 0) \psi_{R,i }^\dagger \psi_{L,i  } (\tau, x) \rangle \cr\cr &\sim&  C^{\mathrm{II}}(\vect{x}) = \frac{\tr(c^\dagger_{i,\alpha}(0) c_{i,\alpha} (\vect{x}) \hat{\rho}^D_c c^\dagger_{i,\alpha}(\vect{x})  c_{i,\alpha}(0) \hat{\rho}_\Omega)}{\tr(\hat{\rho}_c\hat{\rho}_\Omega)}
\eeqn
where $i=1,2,3,4$ and $\alpha$ labels the Dirac index. $\vect{x} = (x, y)$ labels the $2d$ spatial coordinate before the space-time rotation. 

\section{Summary and Discussion}

In this work we investigate the fidelity between the mixed state density matrix of a decohered Chern insulator, and a pure state trivial insulator. We demonstrate that the fidelity is mapped to the partition function of a $(2+0)d $ or $(1+1)D$ CFT, and the R\'{e}nyi relative entropy is mapped to the free energy of the CFT. The decoherence channel can lead to a variety of CFTs with different central charges. We also devised a procedure to extract the central charges, without having to know the details of the order parameters that would lead to nontrivial strange correlators. \cx{We also stress here that the decoherence or measurements caused by the quantum channel is supposed to be weak, and our study focuses on the fate of the fidelity $\cZ$ under weak-measurement; a strong projective measurement on (for example) the fermion density would drive the system into a product state. }

The hermiticity of the density matrix demands that, after space-time rotation, the $(1+1)D$ theory is invariant under transformation $\psi_{L} \rightarrow \psi^\dagger_{R}$, $\psi_{R} \rightarrow \psi^\dagger_{L}$. This alone does not guarantee that the effective $(1+1)D$ CFT be unitary. Indeed, the strange correlator has been utilized as a tool to engineer nonunitary CFTs~\cite{nonunitarySC2,nonunitarySC1,bistrange}. In our current work, the symmetries of the system and the decoherence channel ensure that at least there is no relevant nonhermitian terms in the effective $(1+1)D$ theory. 

As we have seen in this work, the effect of decoherence is mapped to interaction terms in the fidelity, which corresponds to the partition function of the $(1+1)D$ CFT. The same physical picture also applies to all other free fermion topological insulators and topological superconductors (TSC). But as we have mentioned, the decoherence only realizes certain restricted forms of interactions. We leave a complete discussion of TIs and TSCs under decoherence for a future study. 

The authors are supported by the Simons Foundation through the Simons Investigator program. \nmj{The authors also thank Chao-Ming Jian for many helpful discussions and are grateful to Ryan Lanzetta for drawing their attention to Ref. \cite{Casimirbook}.}

\bibliography{CI_measure}

\end{document}